\documentstyle[prc,aps,eqsecnum]{revtex}

\title{On the mass correction of heavy hadrons of arbitray spin
in heavy hadron effective theory
\thanks{Research supported by the
 National Science Council,R.O.C.}}

\author{Tsung-Wen Yeh
and Chien-er Lee}
\address{Department of Physics, National Cheng Kung University\\
Tainan,Taiwan 701, Republic of China\\email: twy@ibm65.phys.ncku.edu.tw
}

\date{NCKU-HEP/95-03  \today}

\tighten
\draft

\begin{document}

\maketitle

\begin{abstract}
{ The mass correction forms of the arbitrary spin heavy
hadrons are derived by using the projection operator method.
The Bjorken sum rule for finite mass is derived by using the results of here .
}
\end{abstract}


\section{Introduction}
\label{sec:Intro}
The Heavy Quark Effective Theory(HQET) has been used to describe
the heavy hadron phenomena successfully in the past few years \cite{HQET}. The
mass correction for HQET is needed to compensate the infinite heavy
quark mass limit simplicity \cite{MassHQET}. In the quark language,
 the heavy  hadrons may be taken as composed of one heavy quark and
 light brown muck. The comfinement interactions between heavy quark
 and light brown muck are almost the QCD energe scale $\Lambda_{QCD}$.
As the mass $m_Q$ of the heavy quark being much larger than $\Lambda_{QCD}$,
 one can specify the ratio of the total momentum $P_{Q}^{\mu}$ over $m_Q$
to be the velocity $v^{\mu}$ which characterize the heavy quark states.
 Since the most momentum of heavy hadron
 is carried by the heavy quark , thus one can identify heavy hadron
total momentum $P_{M}^{\mu}$  to be $Mv^{\mu}$ if the hadron mass $M$ is
equal to $m_Q+\Lambda_H$, where the $\Lambda_H$ is the relevant parameter to
denote the contribution of light brown muck to heavy hadron mass.
After extracting the velocity part of total momentum , the rest
is defined as the residual momentum $k^{\mu}$ of the heavy quark
to denode the momentum fluctation of heavy quark due to interactions with
light brown muck. The systematic expansion in HQET is the expansion in terms
of $k^{\mu}$ and $m_Q^{-1}$. Many mass expansion schemes have been derived
according to different physical and mathematical reasons \cite{masscheme}.
Although full physical green functions are independent of mass expansion
scheme, but the results from HQET are really dependent on mass expansion
scheme. This dependence comes from that one can not calculate to all orders
of mass expansion. Such limitation confronts us to need a good mass expansion
scheme which can at least satisfy the properties of a free field theory
\cite{Projmass}. Mass expansion scheme can be separated into two types, one
is the lagrangian type mass expansion and the other is the field type mass
expansion. The former is derived by functional integral method and the latter
is to solve the mass expansion equation. The idea is to find the transformation
between mass correction field and effective field. Since most physical
interests are about the green functions which are composed of relevant fields.
Thus the field type mass expansion scheme is appropriate. We developed the
projection operator method to meet the requirements of the field type mass
expansion \cite{Projmass}.

If we set the relevant energe scale as the chiral symmetry energe scale
$\Lambda_{\chi}$, then we can combine the heavy quark symmetry and chiral
symmetry to write down the heavy hadron effective theory (HHET)\cite{HHET}.
Such a approach opens an new era for applying the heavy quark symmetry to
the heavy hadron transitions involving light pions. Although mass correction
for HHET is analogous to that of HQET, but the mass expansion scheme of HHET is
still not developed plentily as compared to that of HQET \cite{massHMET}.
The main obstacle is beacuse now the basic objects are heavy hadrons which
are more complicated than heavy quark. To directly apply the mass expansion
scheme in HQET for HHET is not so easy and in some cases will lead to wrong
\cite{massHMET}. In
\cite{pmassHMET} we used the idea of the projection operator method in HQET to
develope the heavy meson effective theory
HMET type projection operator method and get the mass correction.
Here we will extend such a approach to all kinds of heavy hadron. In the
following we will firstly review the projection operator method in HQET
and HMET respectively.

\subsection{Review of projection operator method in HQET}
\label{sec:mhqet}

For a heavy quark field $Q(x)$ we define the correction filed as
\begin{equation}
H(x) = e^{im_Q v\cdot x}Q(x),\label{Hdef}
\end{equation}
\noindent
and the effective field as
\begin{equation}
h_v(x) = \lim_{m_Q \to \infty}H(x),\label{hdef}
\end{equation}
\noindent
 and their Fourier transformation forms as $H,h_v$. In momentum space
$H$ and $h_v$ have their projection operators
\begin{eqnarray}
\Lambda^{+} &=& \sum_{r}H_r\overline{H_r} \nonumber \\
            &=& \frac{1+\not\!v}{2}+\frac{\not\!k}{2m_Q} \label{prjdef1}\\
\Lambda^{+}_v &=& \sum_{r}h_{v,r}\overline{h_{v,r}} \nonumber \\
              &=& \frac{1+\not\!v}{2}  .\label{prjdef2}
\end{eqnarray}
\noindent
By applying $\Lambda^{+}$ on $H$ and noting that the $(1+\not\!v)/2$ part of
$H$ is not equal to $h_v$ , one can have the following mass expansion equation
\begin{eqnarray}
\frac{1-\not\!v}{2}H &=& \frac{\not\!k}{2m_Q}H \nonumber \\
  &=& \frac{\not\! k}{2m_Q - \not\!k}\frac{1+\not\!v}{2}H .\label{masseqn}
\end{eqnarray}
\noindent
Making an ansarz that $((1+\not\!v)/2)H$ is proportional to $h_v$, we can solve
Eq. \ref{masseqn} and get the relation between $H$ and $h_v$ as
\begin{eqnarray}
H = \sqrt{\frac{1+{\not\! k}/{2m_Q}}{1-{\not\! k}/{2m_Q}}}h_v .
\label{Htoh}
\end{eqnarray}

\subsection{Review of projection operator method in HMET}
\label{sec:mhmet}
For heavy mesons with quantum number $J^P = ( 0^{-},1^{-})$ , we can combine
them into a compact heavy meson field $M(x)$
\begin{equation}
M(x) = \frac{m_M + i {\not\! d}}{2m_Q}(-P(x)\gamma_5 +
P^{*\mu}(x)\gamma_{\mu}) . \label{chm}
 \end{equation}
 Define the correction compact heavy meson field as
\begin{eqnarray}
\widehat{M}(x) &=& e^{im_M v \cdot x} M(x) \label{cchm1} \nonumber \\
&=& \left ( \frac{1 + \not\!v}{2} + \frac{i\not\!d}{2m_Q}
+ \frac{\Lambda_M}{m_Q} \right ) (-\widehat{P}(x)\gamma_5 +
\widehat{P}^{*\mu}(x)\gamma_{\mu})
,\label{cchm2}
\end{eqnarray}
\noindent
and the effective compact heavy meson field as
\begin{equation}
H_M(x) = \lim_{m_Q \to \infty} \widehat{M}(x) .\label{echm}
\end{equation}
\noindent
{}From Eq. \ref{cchm2}, we can take pseudoscalar $\widehat{P}(x)$ as example to
identify
\[
\left ( \frac{1 + \not\!v}{2} + \frac{i\not\!d}{2m_Q} \right ) \widehat{P}(x)
\]
\noindent
as its heavy quark part $P_Q(x)$. Thus going to the momentum space and finding
the relevant projection operators, we can do the similar procedure in
subsection.A to get the result
\begin{equation}
\widehat{P} = \sqrt{\frac{1+{\not\! k}/{2m_Q}}{1-{\not\! k}/{2m_Q}}} P_h
\label{Ptoph}
\end{equation}
\noindent
where $\widehat{P}$ is the correction pseudoscalar and $P_h$ is the
$(1+\not\!v)/2$ projected effective pseudoscalar $P_v$. Finally, we combine
everything to arrive at the correction compact heavy meson field related to
effective heavy meson field with the following form
\begin{equation}
\widehat{M} = \left ( 1 + \frac{\Lambda_M}{m_Q} \frac{1 + \not\! v}{2}
  \right ) \sqrt{\frac{1+{\not\! k}/{2m_Q}}{1-{\not\! k}/{2m_Q}}} H_M  .
  \label{MtoHM}
\end{equation}
\noindent
where $H_M$ is defined as
\begin{equation}
H_M = \frac{1+\not\!v}{2}(-P_v\gamma_5+P^{*\mu}_v\gamma_{\mu}) .
\label{HMdef}
\end{equation}

The above review tells us that the heavy quark mass correction for heavy quark
and heavy mesons is irrelevant to the properties of the light brown muck
excepting for the total wave function mass corrections(see below).
Such a superising result can be tracked back to the factorizability of
matrix elements of a heavy quark current between hadronic states into
heavy and light matrix elements in the heavy quark mass infinite limit as
\begin{eqnarray}
 \mathopen{\langle \Psi(v^{\prime})\,|}J(q)\mathclose{|\,\Psi(v) \rangle}
 = \mathopen{\langle Q^{\prime}(v^{\prime}),\,\pm
\frac{1}{2}\,|}J(q)\mathclose{|\,Q(v)\,\pm \frac{1}{2} \rangle}
\mathopen{\langle light,\,v^{\prime},\,j^{\prime},\,m^{\prime_j}
\,|}\mathclose{light,\,v,\,j,\,m_j \rangle }
\label{factr}
\end{eqnarray}
\noindent
Our organization is as following. In  section
\ref{sec:hbaryon} we derive the mass correction for heavy baryons. In section
\ref{sec:hmeson} we derive the mass correction for heavy mesons. In section
\ref{sec:Bjorken} we want to genegraliz the Bjorken sum rule to include the
mass corrections.
In section \ref{sec:sum} is the summary .

\section{Mass correction of heavy baryons}
\label{sec:hbaryon}

To discuss the mass correction of heavy baryon, one need to firstly specify
which quantum number is good or almost good. Since in the infinite heavy
quark mass limit the quantum number of the angular momentum of the light brown
muck , $j$ appears
to be good, this comes from the spin interaction between heavy degree of
freedom (d.o.f) and light
brown muck(l.b.m) is proportional to $1/m_Q$. Applying the same argument for
finit mass cases , one can still use $j$ (almost good)
to classify the finit mass heavy baryons to undo the mass
correction derivation by requiring that the heavy d.o.f is on-shell. The
reason can be thought as following. Although now $j$ is nomore a good one
, but since the heavy d.o.f is on-shell the spin interactions between
heavy d.o.f and l.b.m will not affect the heavy d.o.f's equation of
motion. This means that the complicate interactions should be realized via
the mass expansion series and while one derives the mass correction formalism
 unnecessary to consider these complicate dynamical effects and should leave
them to the lagrangian one gets. So the whole program of the derivation
of the mass correction is now easy and can be picture as that the mass
correction is a transformation between the finite mass field and infinite mass
field ,while both are free. And ,since the on-sell is always true for infinite
field , then the finit mass field should be also on-shell.
The on-shell requirment of heavy d.o.f is the source of the \lq\lq velocity
reparametrization invarint". Thus the whole argument is self-consistent.
The general heavy baryons can be still classified according to the light brown
muck total angular momentum $j\geq 1$ with two specified heavy quark spin
states. Thus we classify the heavy baryons with total angular momentum
$J=j\pm\frac{1}{2}$ and write them as \cite{Book1}
\begin{equation}
J=j+\frac{1}{2}: \,\,\,
 B^{*\mu_1\mu_2\cdots\mu_j}(x) \label{Bj+}
\end{equation}
\begin{equation}
J=j-\frac{1}{2}:  \,\,\,
\mathop{\sum}_{i=1}^{j}\sqrt{\frac{1}{j(2j+1)}}
(\gamma^{\mu_i}+\frac{id^{\mu_i}}{m_{B_j}})\gamma_5
B^{\mu_1\cdots\widehat{\mu}_i\cdots\mu_j}(x) \label{Bj-}
\end{equation}
and the superfield
\begin{eqnarray}
S^{\mu_1\mu_2\cdots\mu_j}(x) =
B^{*\mu_1\mu_2\cdots\mu_j}(x) +
\mathop{\sum}_{i=1}^{j}\sqrt{\frac{1}{j(2j+1)}}
(\gamma^{\mu_i}+\frac{id^{\mu_i}}{m_{B_j}})
\gamma_5B^{\mu_1\cdots\widehat{\mu}_i\cdots\mu_j}(x).
\label{sBj}
\end{eqnarray}
where $\widehat{\mu}_j$ means that this index is subtracted.
These baryons and superfield satisfy their respective equations of motion and
constrains
as following
\begin{eqnarray}
{i\not\!d}B^{*\mu_1\mu_2\cdots\mu_j}(x)&=&
m_{B_j}B^{*\mu_1\mu_2\cdots\mu_j}(x)  ,\label{Bj+eq1} \\
id_{\mu_1}B^{*\mu_1\mu_2\cdots\mu_j}(x)&=&0  ,\label{Bj+eq2} \\
B^{*\mu_1\mu_1\cdots\mu_j}(x)&=&0 ,\label{Bj+eq3}\\
\gamma_{\mu_1}B^{*\mu_1\mu_2\cdots\mu_j}(x)&=&0 ,\label{Bj+eq4}
\end{eqnarray}
\begin{eqnarray}
{i\not\!d}B^{\mu_1\mu_2\cdots\widehat{\mu}_i\mu_j}(x)
&=&m_{B_j}B^{\mu_1\mu_2\cdots\widehat{\mu}_i\mu_j}(x)  ,\label{Bj-eq1} \\
id_{\mu_1}B^{\mu_1\mu_2\cdots\widehat{\mu}_i\mu_j}(x)&=&0  ,\label{Bj-eq2} \\
B^{\mu_1\mu_1\cdots\widehat{\mu}_i\mu_j}(x)&=&0 ,\label{Bj-eq3}
\end{eqnarray}
and
\begin{eqnarray}
{i\not\!d}S^{\mu_1\mu_2\cdots\mu_j}(x)&=&
m_{B_j}S^{\mu_1\mu_2\cdots\mu_j}(x)  ,\label{sBjeq1} \\
id_{\mu_1}S^{\mu_1\mu_2\cdots\mu_j}(x)&=&0  ,\label{sBjeq2} \\
S^{\mu_1\mu_1\cdots\mu_j}(x)&=&0 ,\label{sBjeq3}\\
\gamma_{\mu_1}S^{\mu_1\mu_2\cdots\mu_j}(x)&=&0 .\label{sBjeq4}
\end{eqnarray}
Now we can define the correction heavy baryons according to the above basis
\begin{equation}
{\widehat B}^{*\mu_1\mu_2\cdots\mu_j}(x) =
e^{im_{B_j}v\cdot x}B^{*\mu_1\mu_2\cdots\mu_j}(x) \label{cBj+}
\end{equation}
\begin{equation}
\mathop{\sum}_{i=1}^{j}\sqrt{\frac{1}{j(2j+1)}}(\gamma^{\mu_i}+v^{\mu_i}+
\frac{id^{\mu_i}}{m_{B_j}}) \gamma_5
{\widehat B}^{\mu_1\cdots\widehat{\mu}_i\cdots\mu_j}(x) \\
 =\mathop{\sum}_{i=1}^{j}\sqrt{\frac{1}{j(2j+1)}}(\gamma^{\mu_i}+v^{\mu_i}+
\frac{id^{\mu_i}}{m_{B_j}})
\gamma_5e^{im_{B_j}v\cdot x}B^{\mu_1\cdots\widehat{\mu}_i\cdots\mu_j}(x)
\label{cBj-},
\end{equation}
\begin{equation}
{\widehat{S}^{\mu_1\mu_2\cdots\mu_j}}(x)=
e^{im_{B_j}v\cdot x}B^{*\mu_1\mu_2\cdots\mu_j}(x) +
 \mathop{\sum}_{i=1}^{j}\sqrt{\frac{1}{j(2j+1)}}(\gamma^{\mu_i}+v^{\mu_i}+
\frac{id^{\mu_i}}{m_{B_j}})
\gamma_5e^{im_{B_j}v\cdot x}B^{\mu_1\cdots\widehat{\mu}_i\cdots\mu_j}(x)
\label{scBj}.
\end{equation}

The equations of motion and constrains then become
\begin{eqnarray}
({\not\!v}+\frac{i\not\!d}{m_{B_j}}){\widehat B}^{*\mu_1\mu_2\cdots\mu_j}(x)&=&
{\widehat B}^{*\mu_1\mu_2\cdots\mu_j}(x) ,\label{cBj+eq1} \\
(v_{\mu_1}+\frac{id{\mu_1}}{m_{B_j}})
{\widehat B}^{*\mu_1\mu_2\cdots\mu_j}(x)&=&0  ,\label{cBj+eq2} \\
{\widehat B}^{*\mu_1\mu_1\cdots\mu_j}(x)&=&0 ,\label{cBj+eq3}\\
\gamma_{\mu1}{\widehat B}^{*\mu_1\mu_2\cdots\mu_j}(x)&=&0 ,\label{cBj+eq4}
\end{eqnarray}
\begin{eqnarray}
({\not\!v}+\frac{i\not\!d}{m_{B_j}})
{\widehat B}^{\mu_1\mu_2\cdots\widehat{\mu}_i\mu_j}(x)&=&
{\widehat B}^{\mu_1\mu_2\cdots\widehat{\mu}_i\mu_j}(x),\label{cBj-eq1} \\
(v_{\mu_1}+\frac{id{\mu_1}}{m_{B_j}})
{\widehat B}^{\mu_1\mu_2\cdots\widehat{\mu}_i\mu_j}(x)&=&0 ,\label{cBj-eq2} \\
{\widehat B}^{\mu_1\mu_1\cdots\widehat{\mu}_i\mu_j}(x)&=&0 .\label{cBj-eq3}
\end{eqnarray}
and
\begin{eqnarray}
({\not\!v}+\frac{i\not\!d}{m_{B_j}}){\widehat S}^{\mu_1\mu_2\cdots\mu_j}(x)&=&
{\widehat S}^{*\mu_1\mu_2\cdots\mu_j}(x) ,\label{scBjeq1} \\
(v_{\mu_1}+\frac{id_{\mu_1}}{m_{B_j}})
{\widehat S}^{\mu_1\mu_2\cdots\mu_j}(x)&=&0  ,\label{scBjeq2} \\
{\widehat S}^{\mu_1\mu_1\cdots\mu_j}(x)&=&0 ,\label{scBjeq3}\\
\gamma_{\mu1}{\widehat S}^{\mu_1\mu_2\cdots\mu_j}(x)&=&0 .\label{scBjeq4}
\end{eqnarray}

To define the effective heavy baryons, we firstly separate the mass $m_{B_j}$
into $m_Q + \Lambda_{B_j}$ and apply the limit $m_Q\to\infty$ to obtain
\begin{equation}
B^{*\mu_1\mu_2\cdots\mu_j}_v(x) = \lim_{m_Q\to\infty}
{\widehat B}^{*\mu_1\mu_2\cdots\mu_j}(x), \label{eBj+}
\end{equation}
\begin{equation}
\mathop{\sum}_{i=1}^{j}\sqrt{\frac{1}{j(2j+1)}}(\gamma^{\mu_i}+v^{\mu_i})
\gamma_5
B^{\mu_1\cdots\widehat{\mu}_i\cdots\mu_j}_v(x)
 \\
=\mathop{\sum}_{i=1}^{j}\sqrt{\frac{1}{j(2j+1)}}(\gamma^{\mu_i}+v^{\mu_i})
\gamma_5\lim_{m_Q\to\infty}
\widehat{B}^{\mu_1\cdots\widehat{\mu}_i\cdots\mu_j}(x)  \label{eBj-}.
\end{equation}
\begin{equation}
S^{\mu_1\mu_2\cdots\mu_j}_v(x)
={\lim_{m_Q\to\infty}}{\widehat S}^{\mu_1\mu_2\cdots\mu_j}(x)
,\label{esBj1}
\end{equation}
\begin{equation}
= B^{*\mu_1\mu_2\cdots\mu_j}_v(x) +
\mathop{\sum}_{i=1}^{j}\sqrt{\frac{1}{j(2j+1)}}(\gamma^{\mu_i}+v^{\mu_i})
\gamma_5B^{\mu_1\cdots\widehat{\mu}_i\cdots\mu_j}_v(x)
.\label{esBj2}
\end{equation}

One sees that the effective heavy baryons just those defined in \cite{arbh} and
their related equations of motion and constraints also just defined in there.

Let's concentrate on the derivation of explicit forms of
${\widehat B}^{*\mu_1\mu_2\cdots\mu_j}(x)$ and
$\widehat{B}^{\mu_1\cdots\widehat{\mu}_i\cdots\mu_j}(x)$.
It is better to the momentum space. Apply the heavy quark projection operator
$\Lambda^{+}= ((1+ \not\!v)/2 + {\not\! k}/{2m_Q})$ on
${\widehat B}^{*\mu_1\mu_2\cdots\mu_j}$ and
$\widehat{B}^{\mu_1\cdots\widehat{\mu}_i\cdots\mu_j}$ to project out their
heavy quark contents
\begin{eqnarray}
\widehat{B}^{*\mu_1\mu_2\cdots\mu_j}_Q &:=&
(\frac{1+ \not\!v}{2} + \frac{\not\! k}{2m_Q})
\widehat{B}^{*\mu_1\mu_2\cdots\mu_j} ,\label{QcBj+} \\
\widehat{B}^{\mu_1\cdots\widehat{\mu}_i\cdots\mu_j}_Q &:=&
(\frac{1+ \not\!v}{2} + \frac{\not\! k}{2m_Q})
\widehat{B}^{\mu_1\cdots\widehat{\mu}_i\cdots\mu_j} .\label{QcBj-}
\end{eqnarray}
Following the same procedure in subsection \ref{sec:mhmet}, one can finally
get
\begin{eqnarray}
\widehat{B}^{*\mu_1\mu_2\cdots\mu_j}_Q &=&
\sqrt{\frac{1+{\not\! k}/{2m_Q}}{1-{\not\! k}/{2m_Q}}}
B^{*\mu_1\mu_2\cdots\mu_j}_{v,h} ,\label{QmcBj+} \\
\widehat{B}^{\mu_1\cdots\widehat{\mu}_i\cdots\mu_j}_Q &=&
\sqrt{\frac{1+{\not\! k}/{2m_Q}}{1-{\not\! k}/{2m_Q}}}
B^{\mu_1\cdots\widehat{\mu}_i\cdots\mu_j}_{v,h} ,\label{QmcBj-}
\end{eqnarray}
and use the following relations
\begin{eqnarray}
\left [ \frac{1 + \not\! v }{2},
\sqrt{\frac{1+{\not\! k}/{2m_Q}}{1-{\not\! k}/{2m_Q}}} \right ]
&=& -\frac{\not\! k}{2m_Q}
\sqrt{\frac{1+{\not\! k}/{2m_Q}}{1-{\not\! k}/{2m_Q}}} ,\label{rel1} \\
\sqrt{\frac{1+{\not\! k}/{2m_Q}}{1-{\not\! k}/{2m_Q}}}
\left ( \frac{1 + \not\! v }{2} \right )
\left\{
\begin{array}{ll}
 B^{*\mu_1\mu_2\cdots\mu_j}_v   \\
 B^{\mu_1\cdots\widehat{\mu}_i\cdots\mu_j}_v
\end{array}\right.
&=&\left ( \frac{1+ \not\!v}{2} + \frac{\not\! k}{2m_Q} \right )
\sqrt{\frac{1+{\not\! k}/{2m_Q}}{1-{\not\! k}/{2m_Q}}}
\left\{
\begin{array}{ll}
 B^{*\mu_1\mu_2\cdots\mu_j}_v  \\
 B^{\mu_1\cdots\widehat{\mu}_i\cdots\mu_j}_v .
\end{array}\right. \label{rel2}
\end{eqnarray}

 So the results we get
\begin{eqnarray}
\widehat{B}^{*\mu_1\mu_2\cdots\mu_j} &=&
\sqrt{\frac{1+{\not\! k}/{2m_Q}}{1-{\not\! k}/{2m_Q}}}
B^{*\mu_1\mu_2\cdots\mu_j}_v ,\label{ceBj+1} \\
\widehat{B}^{\mu_1\cdots\widehat{\mu}_i\cdots\mu_j} &=&
\sqrt{\frac{1+{\not\! k}/{2m_Q}}{1-{\not\! k}/{2m_Q}}}
B^{\mu_1\cdots\widehat{\mu}_i\cdots\mu_j}_v ,\label{ceBj-1}
\end{eqnarray}
and the corresponding $J=j\pm\frac{1}{2}$ baryons are
\begin{equation}
\widehat{B}^{*\mu_1\mu_2\cdots\mu_j} =
\sqrt{\frac{1+{\not\! k}/{2m_Q}}{1-{\not\! k}/{2m_Q}}}
B^{*\mu_1\mu_2\cdots\mu_j}_v \label{ceBj+2},
\end{equation}
\begin{equation}
\sum_{i=1}^{j}\sqrt{\frac{1}{j(2j+1)}}(\gamma^{\mu_i}+v^{\mu_i}
+\frac{k^{\mu_i}}{m_{B_j}})\gamma_5
\widehat{B}^{\mu_1\cdots\widehat{\mu}_i\cdots\mu_j}
=\sum_{i=1}^{j}\sqrt{\frac{1}{j(2j+1)}}(\gamma^{\mu_i}+v^{\mu_i}
+\frac{k^{\mu_i}}{m_{B_j}})  \gamma_5
\sqrt{\frac{1+{\not\! k}/{2m_Q}}{1-{\not\! k}/{2m_Q}}}
B^{\mu_1\cdots\widehat{\mu}_i\cdots\mu_j}_v \label{ceBj-2},
\end{equation}
and the superfield
\begin{eqnarray}
\widehat{S}^{\mu_1\mu_2\cdots\mu_j}
=\sqrt{\frac{1+{\not\! k}/{2m_Q}}{1-{\not\! k}/{2m_Q}}}
B^{*\mu_1\mu_2\cdots\mu_j}_v +
\sum_{i=1}^{j}\sqrt{\frac{1}{j(2j+1)}}(\gamma^{\mu_i}+v^{\mu_i}
+\frac{k^{\mu_i}}{m_{B_j}})  \gamma_5
\sqrt{\frac{1+{\not\! k}/{2m_Q}}{1-{\not\! k}/{2m_Q}}}
B^{\mu_1\cdots\widehat{\mu}_i\cdots\mu_j}_v
\label{cesBj}.
\end{eqnarray}

For the $j = 0$ we just list the result
\begin{equation}
\widehat{B} =\sqrt{\frac{1+{\not\! k}/{2m_Q}}{1-{\not\! k}/{2m_Q}}}
B_v
\end{equation}
Having arrived at these mass correction heavy baryons, we may see that the
whole mass correction factor
\[
\sqrt{\frac{1+{\not\! k}/{2m_Q}}{1-{\not\! k}/{2m_Q}}}
\]
is very similar to the multiplying factor for the wave function. The reader
may check that this factor can be written as
\[
\exp\left({\frac{1}{2}\ln{\frac{1+{\not\! k}/{2m_Q}}{1-{\not\! k}/{2m_Q}}}}
\right ).
\]
It can have its inverse form under Dirac conjugate
\[
\exp\left(-{\frac{1}{2}\ln{\frac{1+{\not\! k}/{2m_Q}}{1-{\not\! k}/{2m_Q}}}}
\right ).
\]
So,it is easy to see that such factor is unitary. This is an important
property to make sure that the unitarity of whole theory will not be broken
by mass correction.

\section{Mass correction of heavy mesons}
\label{sec:hmeson}
In this sction we want to discuss the mass correction of heavy mesons. The
light brown muck for heavy meson is more complicated than that in heavy
baryons. The light brown muck in heavy meson may appear in two kinds of
states specified according to $j = l \pm\frac12,l \geq 1$. We identify these
states as
\begin{equation}
j = l + \frac12: \,\,\,
\overline{R}^{\mu_1\cdots\mu_l}(x),
\end{equation}
\begin{equation}
j = l - \frac12: \,\,\,
\gamma_5\overline{R}^{\mu_1\cdots\mu_l}(x).
\end{equation}
They will satisfy the following equations of motion and constraints.
For $j = l + \frac12$ states are
\begin{eqnarray}
\overline{R}^{\mu_1\cdots\mu_l}(x)i\overleftarrow{\not\!d}&=&0 , \\
\overline{R}^{\mu_1\cdots\mu_l}(x)\overleftarrow{id_{\mu1}}&=&0, \\
\overline{R}^{\mu_1\mu_1\cdots\mu_l}(x)&=&0,\\
\overline{R}^{\mu_1\cdots\mu_l}(x)\gamma_{\mu_1}&=&0,
\end{eqnarray}
and for $j = l - \frac12$ states are
\begin{eqnarray}
\overline{R}^{\mu_1\cdots\mu_l}(x)i\overleftarrow{\not\!d}\gamma_5&=&0 , \\
\overline{R}^{\mu_1\cdots\mu_l}(x)i\overleftarrow{d_{\mu1}}&\gamma_5=&0, \\
\overline{R}^{\mu_1\mu_1\cdots\mu_l}(x)\gamma_5&=&0,\\
\overline{R}^{\mu_1\cdots\mu_l}(x)\gamma_5\gamma_{\mu_1}&=&0.
\end{eqnarray}
If we redefine them as
\begin{equation}
\widehat{\overline{R}}^{\mu_1\cdots\mu_l}(x) :=
\overline{R}^{\mu_1\cdots\mu_l}(x)e^{i\Lambda_{M_j}v\cdot x},
\end{equation}
\begin{equation}
\widehat{\overline{R}}^{\mu_1\cdots\mu_l}(x)\gamma_5 :=
\overline{R}^{\mu_1\cdots\mu_l}(x)\gamma_5e^{i\Lambda_{M_j}v\cdot x}.
\end{equation}
, then their equations of motion and constrains become
\begin{eqnarray}
\widehat{\overline{R}}^{\mu_1\cdots\mu_l}(x)
(\not\!v+\frac{i\overleftarrow{\not\!d}}{\Lambda_{M_j}})&=&0, \\
\widehat{\overline{R}}^{\mu_1\cdots\mu_l}(x)
(v_{\mu1}+\frac{i\overleftarrow{d_{\mu1}}}{\Lambda_{M_j}})&=&0, \\
\widehat{\overline{R}}^{\mu_1\mu_1\cdots\mu_l}(x)&=&0,\\
\widehat{\overline{R}}^{\mu_1\cdots\mu_l}(x)\gamma_{\mu_1}&=&0,
\end{eqnarray}
and
\begin{eqnarray}
\widehat{\overline{R}}^{\mu_1\cdots\mu_l}(x)\gamma_5
(\not\!v+\frac{i\overleftarrow{\not\!d}}{\Lambda_{M_j}})&=&0 , \\
\widehat{\overline{R}}^{\mu_1\cdots\mu_l}(x)\gamma_5
(v_{\mu1}+\frac{i\overleftarrow{d_{\mu1}}}{\Lambda_{M_j}})&=&0, \\
\widehat{\overline{R}}^{\mu_1\mu_1\cdots\mu_l}(x)\gamma_5&=&0,\\
\widehat{\overline{R}}^{\mu_1\cdots\mu_l}(x)\gamma_5\gamma_{\mu_1}&=&0.
\end{eqnarray}
The additional limit accompanied with heavy quark mass infinite limit
\[
\frac{i\not\!d}{\Lambda_{M_j}}\rightarrow1
\]
will show that the results are just those defined in \cite{arbh}.So, we apply
such a limit on the redefined light brown muck
\begin{equation}
\overline{R}^{\mu_1\cdots\mu_l}_v(x) =
\lim_{{{i\not d}/{\Lambda_{M_j}}\rightarrow1}
\atop{m_Q\to\infty}}
\widehat{\overline{R}}^{\mu_1\cdots\mu_l}(x),
\end{equation}
\begin{equation}
\overline{R}^{\mu_1\cdots\mu_l}_v(x)\gamma_5 =
\lim_{{{i\not d}/{\Lambda_{M_j}}\rightarrow1}
\atop{m_Q\to\infty}}
\widehat{\overline{R}}^{\mu_1\cdots\mu_l}(x)\gamma_5.
\end{equation}
Having these basis, we write our compact heavy mesons according to the above
two kinds of light brown muck
\[
J^P = (l^{-},(l+1)^{-})_{j = l + 1/2}:   \,\,\,
M^{(-)\mu_1\cdots\mu_l}(x) = \]
\begin{equation}
\left ( \frac{m_{M_j} + i\not\!d}{2m_Q} \right )
\left ( \sqrt{\frac{2l+1}{2l+2}}P^{\nu_1\cdots\nu_l}(x)\gamma_5
\left [ \delta^{\mu_1\cdots\mu_l}_{\nu_1\cdots\nu_l}
-\frac{1}{2j+1}\sum_{i=1}^{l}\gamma_{\nu_i}
(\gamma^{\mu_i}-\frac{i\overleftarrow{d^{\mu_i}}}{m_{M_j}})
\delta^{\mu_1\cdots\widehat{\mu}_i\cdots\mu_l}_
{\nu_1\cdots\widehat{\nu}_i\cdots\nu_l} \right ] +
P^{*\mu_1\cdots\mu_{l+1}}(x)\gamma_{\mu_{l+1}} \right ), \label{Mj+}
\end{equation}
and
\[
J^P = ((l-1)^{+},(l)^{+})_{j = l - 1/2}: \,\,\,
M^{(+)\mu_1\cdots\mu_l}(x) = \]
\begin{equation}
\left ( \frac{m_{M_j} + i\not\!d}{2m_Q} \right )
\left ( \sqrt{\frac{2l+1}{2l+2}}S^{\nu_1\cdots\nu_l}(x)
\left [ \delta^{\mu_1\cdots\mu_l}_{\nu_1\cdots\nu_l}
-\frac{1}{2j+1}\sum_{i=1}^{l}\gamma_{\nu_i}
(\gamma^{\mu_i}+\frac{i\overleftarrow{d^{\mu_i}}}{m_{M_j}})
\delta^{\mu_1\cdots\widehat{\mu}_i\cdots\mu_l}_
{\nu_1\cdots\widehat{\nu}_i\cdots\nu_l} \right ] +
S^{*\mu_1\cdots\mu_{l+1}}(x)\gamma_5\gamma_{\mu_{l+1}} \right ). \label{Mj-}
\end{equation}
Redefine these heavy mesons
\[\widehat{M}^{(-)\mu_1\cdots\mu_l}(x) = e^{im_{M_j}v\cdot x}
 M^{(-)\mu_1\cdots\mu_l}(x)  \]
\[
=\left ( \frac{1 + \not\!v}{2}+\frac{i\not\!d}{2m_Q}
+\frac{\Lambda_{M_j}}{m_Q}\frac{1 + \not\!v}{2} \right ) \times
\]
\begin{equation}
\left ( \sqrt{\frac{2l+1}{2l+2}}\widehat{P}^{\nu_1\cdots\nu_l}(x)\gamma_5
\left [ \delta^{\mu_1\cdots\mu_l}_{\nu_1\cdots\nu_l}
-\frac{1}{2j+1}\sum_{i=1}^{l}\gamma_{\nu_i}
(\gamma^{\mu_i}-v^{\mu_i}-\frac{i\overleftarrow{d^{\mu_i}}}{m_{M_j}})
\delta^{\mu_1\cdots\widehat{\mu}_i\cdots\mu_l}_
{\nu_1\cdots\widehat{\nu}_i\cdots\nu_l} \right ] +
\widehat{P}^{*\mu_1\cdots\mu_{l+1}}(x)\gamma_{\mu_{l+1}} \right ),
 \label{cMj+}
\end{equation}
and
\[\widehat{M}^{(+)\mu_1\cdots\mu_l}(x) = e^{im_{M_j}v\cdot x}
M^{(+)\mu_1\cdots\mu_l}(x)  \]
\[
=\left (\frac{1 + \not\!v}{2} \frac{i\not\!d}{2m_Q}
+\frac{\Lambda_{M_j}}{m_Q}\frac{1 + \not\!v}{2} \right ) \times
\]
\begin{equation}
\left ( \sqrt{\frac{2l+1}{2l+2}}\widehat{S}^{\nu_1\cdots\nu_l}(x)
\left [ \delta^{\mu_1\cdots\mu_l}_{\nu_1\cdots\nu_l}
-\frac{1}{2j+1}\sum_{i=1}^{l}\gamma_{\nu_i}
(\gamma^{\mu_i}+v^{\mu_i}+\frac{i\overleftarrow{d^{\mu_i}}}{m_{M_j}})
\delta^{\mu_1\cdots\widehat{\mu}_i\cdots\mu_l}_
{\nu_1\cdots\widehat{\nu}_i\cdots\nu_l} \right ] +
\widehat{S}^{*\mu_1\cdots\mu_{l+1}}(x)\gamma_5\gamma_{\mu_{l+1}} \right ).
\label{cMj-}
\end{equation}
Go to the momentum space and follow the similar prcedure in subsection
\ref{sec:mhmet}. We finally get the mass correction compact heavy mesons
related to the effective compact heavy mesons as
\[
J^P = (l^{-},(l+1)^{-})_{j = l + 1/2}:
\]
\[\widehat{M}^{(-)\mu_1\cdots\mu_l} =
\left ( 1 +\frac{\Lambda_{M_j}}{m_Q}\frac{1 + \not\!v}{2} \right )
\sqrt{\frac{1+{\not\! k}/{2m_Q}}{1-{\not\! k}/{2m_Q}}}
 M^{(-)\mu_1\cdots\mu_l}_v  \]
\[
=\left ( 1 +\frac{\Lambda_{M_j}}{m_Q}\frac{1 + \not\!v}{2} \right )
\sqrt{\frac{1+{\not\! k}/{2m_Q}}{1-{\not\! k}/{2m_Q}}} \times
\]
\begin{equation}
\left (\frac{1 + \not\!v}{2} \right )
\left ( \sqrt{\frac{2l+1}{2l+2}}P^{\nu_1\cdots\nu_l}_v\gamma_5
\left [ \delta^{\mu_1\cdots\mu_l}_{\nu_1\cdots\nu_l}
-\frac{1}{2j+1}\sum_{i=1}^{l}\gamma_{\nu_i}
(\gamma^{\mu_i}-v^{\mu_i}-\frac{k^{\mu_i}}{m_{M_j}})
\delta^{\mu_1\cdots\widehat{\mu}_i\cdots\mu_l}_
{\nu_1\cdots\widehat{\nu}_i\cdots\nu_l} \right ] +
P^{*\mu_1\cdots\mu_{l+1}}_v\gamma_{\mu_{l+1}} \right ), \label{ceMj+}
\end{equation}
and
\[
J^P = ((l-1)^{+},(l)^{+})_{j = l - 1/2}:
\]
\[\widehat{M}^{(+)\mu_1\cdots\mu_l} =
\left (1+\frac{\Lambda_{M_j}}{m_Q}\frac{1 + \not\!v}{2} \right )
\sqrt{\frac{1+{\not\! k}/{2m_Q}}{1-{\not\! k}/{2m_Q}}}
M^{(+)\mu_1\cdots\mu_l}_v  \]
\[
=\left (1+\frac{\Lambda_{M_j}}{m_Q}\frac{1 + \not\!v}{2} \right )
\sqrt{\frac{1+{\not\! k}/{2m_Q}}{1-{\not\! k}/{2m_Q}}} \times
\]
\begin{equation}
\left (\frac{1 + \not\!v}{2} \right )
\left ( \sqrt{\frac{2l+1}{2l+2}}S^{\nu_1\cdots\nu_l}_v
\left [ \delta^{\mu_1\cdots\mu_l}_{\nu_1\cdots\nu_l}
-\frac{1}{2j+1}\sum_{i=1}^{l}\gamma_{\nu_i}
(\gamma^{\mu_i}+v^{\mu_i}+\frac{k^{\mu_i}}{m_{M_j}})
\delta^{\mu_1\cdots\widehat{\mu}_i\cdots\mu_l}_
{\nu_1\cdots\widehat{\nu}_i\cdots\nu_l} \right ] +
S^{*\mu_1\cdots\mu_{l+1}}_v\gamma_5\gamma_{\mu_{l+1}} \right ).
\label{ceMj-}
\end{equation}
For the $j = \pm\frac12,l=0$ heavy mesons can be refered to \ref{sec:mhmet}
. The whole derivation of the mass correction heavy mesons is straightward
and repeates same techniques. If our present heavy quark effective theory
approach is right, then our projection operator methed presented here will
reveal itself to be a good one.

One remark should be noted that the normalization convections for baryons
and mesons are different. The baryons' convection is as usual while that of
mesons is defined as $\sqrt{m_Q}$.
\section{Mass correction of Bjorken sum rule}
\label{sec:Bjorken}
 As an application of our mass correction of heavy hadrons, we choose to
 use the formalisms derived here to the Bjorken sum rule. Firstly, let's
 look at the following situation

\begin{eqnarray}
\sum_{s,s^{\prime}} \mathopen{\langle
h(v,s)\,|}\overline{Q}_v\Gamma
Q^{\prime}_{v^{\prime}}\mathclose{|\,h(v^{\prime},s^{\prime})
\,\rangle}  \mathopen{\langle h(v^{\prime},s^{\prime})\,|}
\overline{Q}^{\prime}_{v^{\prime}}\Gamma Q_v\mathclose{|\,h(v,s) \,\rangle}
\nonumber \\ = \sum_{n^{\prime},s^{\prime}}
\mathopen{\langle \Psi(v)\,
|}\overline{Q}\Gamma Q^{\prime}_{v^{\prime}}
\mathclose{|\,X^{n^\prime}(v^{\prime},s^{\prime})\,\rangle}
\mathopen{\langle X^{n^\prime}(v^{\prime},s^{\prime})
\,|}\overline{Q}^{\prime}_{v^{\prime}}\Gamma Q_v\mathclose{|\,\Psi(v) \,\rangle
},
\label{bjor1}
\end{eqnarray}
where $\overline{Q}_v\Gamma Q^{\prime}_{v^{\prime}}
, \overline{Q}^{\prime}_{v^{\prime}}\Gamma Q_v$ are the the correction heavy
quark currents ,$\Psi(v)$ is the effective baryon $l=0$ state and
$X^{n^\prime}(v^{\prime},s^{\prime})$ are the baryons intermediate states.
Since the infinite mass limit $m_Q,m_{Q^{\prime}}\to\infty$ of the
 Eq. \ref{bjor1} will reproduce the Bjorken sum rule,
 we can use this formalism to study its mass correction formalism.
To use the formalisms derived here , one can use the following matching
identtities
\begin{equation}
\overline{Q}_v \Gamma Q_{v^{\prime}}^{\prime}
\rightarrow C_{B,X(I)}\overline{\widehat{B}}_v
\widehat{\Gamma}_{v,v^{\prime}}
\widehat{X^{n^{\prime}}}_{v^{\prime}},
\label{bjor3}
\end{equation}
if the final and initial states are $\langle \Psi(v) \, |$ and
$| X^{n^{\prime}}(v^{\prime},s^{\prime})\, \rangle$. We substitute these into
Eq. \ref{bjor1} and we obtain
\begin{eqnarray}
\sum_{s,s^{\prime}} \mathopen{\langle
h(v,s)\,|}\overline{Q}_v\Gamma
Q^{\prime}_{v^{\prime}}\mathclose{|\,h(v^{\prime},s^{\prime})
\,\rangle}  \mathopen{\langle h(v^{\prime},s^{\prime})\,|}
\overline{Q}^{\prime}_{v^{\prime}}\Gamma Q_v\mathclose{|\,h(v,s) \,\rangle}
\nonumber \\ = \sum_{n^{\prime},s^{\prime}}
\mathopen{\langle \Psi(v)\,
|}C_{B,X(I)}\overline{\widehat{B}}_v \widehat{\Gamma}_{v,v^{\prime}}
\widehat{X^{n^{\prime}}}_{v^{\prime}}
\mathclose{|\,X^{n^\prime}(v^{\prime},s^{\prime})\,\rangle}
\mathopen{\langle X^{n^\prime}(v^{\prime},s^{\prime})
\,|}C_{X^{\prime}(I^{\prime}),B}
\overline{\widehat{X^{n^{\prime}}}}_{v^{\prime}}
\widehat{\Gamma}_{v^{\prime},v}
\widehat{B}_v \mathclose{|\,\Psi(v) \,\rangle },
\label{bjor4}
\end{eqnarray}
where the $C_{B,X(I)}$ and $C_{X^{\prime}(I^{\prime}),B}$ are relevant
form factors with corresponding
$\widehat{\Gamma}_{v^{\prime},v}, \widehat{\Gamma}_{v^{\prime},v}$ which are
relevant velocity variable,$g_{\mu\nu}$ and $\gamma$-matrix resp..
Thus we have derived the mass correction Bjorken sum rule. To further study
the above equation , we can substitute the mass expansion forms of the relevant
baryon field operators. For the meson case we also have the same results. In
general , if we let $ \widehat{H}_v$ as any correction heavy hadrons , the
Bjorken sum rule for finite mass should take the following form
\begin{eqnarray}
\sum_{s,s^{\prime}} \mathopen{\langle
h(v,s)\,|}\overline{Q}_v\Gamma
Q^{\prime}_{v^{\prime}}\mathclose{|\,h(v^{\prime},s^{\prime})
\,\rangle}  \mathopen{\langle h(v^{\prime},s^{\prime})\,|}
\overline{Q}^{\prime}_{v^{\prime}}\Gamma Q_v\mathclose{|\,h(v,s) \,\rangle}
\nonumber \\ = \sum_{n^{\prime},s^{\prime}}
\mathopen{\langle H(v)\,
|}C_{H,X(I)}\overline{\widehat{H}}_v
\widehat{\Gamma}_{v,v^{\prime}}
\widehat{X^{n^{\prime}}}_{v^{\prime}}
\mathclose{|\,X^{n^\prime}(v^{\prime},s^{\prime})\,\rangle}
\mathopen{\langle X^{n^\prime}(v^{\prime},s^{\prime})
\,|}C_{X^{\prime}(I^{\prime}),H}
\overline{\widehat{X^{n^{\prime}}}}_{v^{\prime}}
\widehat{\Gamma}_{v^{\prime},v}
\widehat{H}_v \mathclose{|\,H(v) \,\rangle }.
\label{bjor5}
\end{eqnarray}
This general formalism then can be studied for every order of
$1/m_{Q(Q^{\prime})}$. The detail structure will need careful analysis of every
related matrix element, so we leave it to our future publication.

\section{Summary}
\label{sec:sum}
We have derived the mass correction of heavy hadrons of arbitrary spin. We list
them by their relevant quantum numbers. These correction field operators can
be used to derive mass correction lagrangian for heavy hadron effective theory.
As an application we use them to derive the Bjorken sum rule for finite mass.

\end{document}